\documentclass[conference,compsoc]{IEEEtran}
\ifCLASSOPTIONcompsoc
  \usepackage[nocompress]{cite}
\else
  \usepackage{cite}
\fi

\ifCLASSINFOpdf
   \usepackage[pdftex]{graphicx}
   \graphicspath{{images/}}
\else
  \usepackage[dvips]{graphicx}
  \graphicspath{{images/}}
\fi
\usepackage[cmex10]{amsmath}
\usepackage{listings}
\usepackage{color}

\usepackage{microtype}
\usepackage{graphicx}
\usepackage{subcaption}
\usepackage{booktabs} % for professional tables

\usepackage{fvextra}
\usepackage{algorithmic}
\usepackage{hyperref}
\usepackage{booktabs}
\usepackage{multirow}
\usepackage[table]{xcolor}
\usepackage{hyperref}

\definecolor{codegreen}{rgb}{0,0.6,0}
\definecolor{codegray}{rgb}{0.5,0.5,0.5}
\definecolor{codepurple}{rgb}{0.58,0,0.82}
\definecolor{backcolour}{rgb}{0.95,0.95,0.92}
 
\lstdefinestyle{mystyle}{
    backgroundcolor=\color{backcolour},   
    commentstyle=\color{codegreen},
    keywordstyle=\color{magenta},
    numberstyle=\tiny\color{codegray},
    stringstyle=\color{codepurple},
    basicstyle=\footnotesize,
    breakatwhitespace=false,         
    breaklines=true,                 
    captionpos=b,                    
    keepspaces=true,                 
    numbers=left,                    
    numbersep=5pt,                  
    showspaces=false,                
    showstringspaces=false,
    showtabs=false,
    tabsize=2
}
 
\begin{document}
%
% paper title
% Titles are generally capitalized except for words such as a, an, and, as,
% at, but, by, for, in, nor, of, on, or, the, to and up, which are usually
% not capitalized unless they are the first or last word of the title.
% Linebreaks \\ can be used within to get better formatting as desired.
% Do not put math or special symbols in the title.
\title{MarkIt: Training-Free Visual Markers for Precise Video Temporal Grounding}

% author names and affiliations
% use a multiple column layout for up to three different
% affiliations
\author{
\IEEEauthorblockN{Pengcheng Fang}
\IEEEauthorblockA{Unviersity of Southampton}
\and
\IEEEauthorblockN{Yuxia Chen}
\IEEEauthorblockA{Chengdu Unviersity of Technology}
\and
\IEEEauthorblockN{Xiaohao Cai}
\IEEEauthorblockA{Unviersity of Southampton}
}
% make the title area
\maketitle

\newcommand{\MarkIt}{\textsc{MarkIt}}
\newcommand{\NumPro}{NumPro}
\newcommand{\VTG}{\textsc{VTG}}
\newcommand{\mIoU}{\text{mIoU}}
\newcommand{\mAP}{\text{mAP}}
\newcommand{\HIT}{\text{HIT@1}}
\newcommand{\R}[1]{\text{R@}#1}
\newcommand{\Render}{\mathrm{Render}}
\newcommand{\Parse}{\mathrm{Parse}}
\newcommand{\QTwoM}{\mathcal{B}} % Q2M-Bridge operator (query-to-mask)

% As a general rule, do not put math, special symbols or citations
% in the abstract
\begin{abstract}
Video temporal grounding (VTG) aims to localize the start and end timestamps of the event described by a given query within an untrimmed video. Despite the strong open-world video understanding and recognition ability of video language large models (Vid-LLMs), outputting precise temporal grounding information remains challenging, since explicit temporal cues are scarce in untrimmed videos, and query-relevant entities are hard to track consistently across the video timeline. In this paper, we present \MarkIt{}, a training-free framework that transforms an input video into a query-conditioned marked video, which empowers Vid-LLMs to generate more reliable temporal localization predictions. The core component of \MarkIt{} is an annotation-free query-to-mask grounding bridge (Q2M-Bridge). Given a natural-language query, it automatically derives a compact set of canonical subject tags through linguistic parsing and normalization, then maps these tags to query-conditioned instance masks using text-conditioned open-vocabulary segmentation. The bridge also embeds lightweight semantic instance markers and a persistent frame index into each frame, effectively transforming long-range temporal reasoning into explicit visual cues for Vid-LLMs. \MarkIt{} adopts an inference-time plug-and-play design, needs no modifications to Vid-LLM weights, and is fully compatible with supervised fine-tuning. Experiments conducted on multiple mainstream moment retrieval and highlight detection benchmarks demonstrate that \MarkIt {} achieves state-of-the-art results, delivering consistent temporal grounding improvements across a wide range of existing models. \href{https://github.com/PengchengFang-cs/MarkIt}{github.com/PengchengFang-cs/MarkIt}.

\end{abstract}

\section{Introduction}
\label{sec:intro}
Video temporal grounding (VTG) is a fundamental task in video understanding that aims to localize the start and end timestamps of an event described by a natural-language query within an untrimmed video. VTG requires models to align linguistic descriptions with temporally coherent visual evidence, which is challenging due to complex dynamics and ambiguous visual cues, demanding joint reasoning over visual semantics and temporal structures.

Recent advances in video language large models (Vid-LLMs) have led to substantial progress in open-world video understanding with strong zero-shot recognition capabilities~\cite{lin2023univtg,liu2021context,chen2024internvl,qu2024chatvtg}. However, despite their success in high-level understanding tasks, Vid-LLMs still struggle to produce precise and reliable temporal localization, often exhibiting temporal drift and hallucinated boundaries.

This difficulty arises from two tightly coupled challenges intrinsic to VTG. First, untrimmed videos lack explicit and precise temporal annotations, requiring models to infer relevant moments without reliable absolute or relative time references~\cite{guo2025vtg,zeng2024timesuite}. Second, queries often involve specific entities or actions that must be consistently identified and tracked over time, despite occlusions, appearance variations, and the coexistence of multiple visually similar instances.~\cite{jung2025consistency} Crucially, these two challenges are interdependent: failures in maintaining visual correspondence directly undermine temporal localization, while ambiguous temporal reasoning further complicates consistent entity tracking.

To address this entanglement, we propose to externalize both temporal reference and visual correspondence directly into the video representation itself. We introduce \MarkIt{}, a training-free framework that transforms an input video into a query-conditioned marked video, where explicit visual cues encode both temporal position and query-relevant entities. By augmenting each frame with persistent frame indices and semantic instance markers, \MarkIt{} converts VTG from an implicit spatio-temporal reasoning problem into one of reading explicit temporal and semantic signals, substantially simplifying the grounding process.

At the core of \MarkIt{} is an annotation-free query-to-mask grounding bridge, termed Q2M-Bridge, which aims to explicitly connect natural-language queries with visual evidence in videos. Given a query, Q2M-Bridge first derives a compact set of canonical subject tags through linguistic parsing and normalization. These tags are then grounded into per-frame instance masks via text-conditioned open-vocabulary segmentation. The resulting instance masks are rendered as lightweight visual overlays with semantic labels and a persistent frame index, providing explicit visual references that enable Vid-LLMs to more effectively associate query semantics with temporally grounded visual content.

\MarkIt{} operates entirely at inference time and does not require any modification to the parameters of Vid-LLMs, making it fully compatible with existing models as well as supervised fine-tuning pipelines. Owing to its plug-and-play design, \MarkIt{} can be readily integrated into a wide range of Vid-LLM architectures. We evaluate \MarkIt{} on multiple mainstream VTG benchmarks, including moment retrieval on Charades-STA and ActivityNet and highlight detection on QVHighlights. Experimental results show that \MarkIt{} achieves state-of-the-art results, delivering consistent temporal grounding improvements across a wide range of existing models (e.g., Figure 1). Our main contributions are summarized as follows:

\begin{itemize}
\setlength{\itemsep}{4pt}
\setlength{\parsep}{4pt}
\setlength{\parskip}{4pt}
    \item We introduce \MarkIt{}, a training-free and plug-and-play input rewriting framework that injects explicit temporal references and query-aware visual correspondences into video inputs, substantially improving VTG without modifying Vid-LLM parameters and remaining compatible with supervised fine-tuning.
    \item We propose a scalable query-to-marker generation pipeline that converts natural-language queries into canonical subject tags and grounds them into per-frame instance masks via text-conditioned open-vocabulary segmentation, enabling robust semantic marker construction under open-world queries.
    \item We evaluate \MarkIt{} on moment retrieval and highlight detection benchmarks under both training-free and supervised settings, achieving state-of-the-art performance across multiple Vid-LLMs.
\end{itemize}

\section{Related Work}

\subsection{Video Temporal Grounding}
VTG aims to temporally localize the video segments in an untrimmed video that correspond to a natural-language query~\cite{xiao2024bridging}. VTG is commonly instantiated in four task settings: video moment retrieval~\cite{fu2025video,li2024llava,xu2025zero,jung2025background},
%\cite{zliu2024bench,fu2025video,li2024llava,xu2025zero,jung2025background}
dense video captioning~\cite{qasim2025dense,estevam2025dense,xiong2024lvd}, video highlight detection~\cite{islam2025unsupervised,xiong2019less,badamdorj2022contrastive,moon2023query}, and temporally grounded video question answering~\cite{lei2020tvqa+,liu2024timecraft,di2024grounded,xiao2024can,wu2025survey}. 

In the past two years, many studies have explored approaches to improve temporal grounding for VTG. One direction fine-tunes Vid-LLMs with temporally annotated instruction prompts, such as timestamps or frame indices, to learn explicit query-to-time alignment~\cite{guo2025tar,wang2025videoitg,li2024towards}. However, such fine-tuning-based solutions may still suffer from limited generalization under standard supervised fine-tuning objectives, motivating complementary strategies that reduce reliance on task-specific adaptation~\cite{wu2025generalization}. 

Beyond supervision, another line augments visual inputs with temporal cues, including timestamp markers, frame-index prompting, or temporal embeddings, enabling localization via conditional reasoning~\cite{wu2025number,huang2024vtimellm,ren2024timechat,guo2025vtg}. Most related to our work, Number It overlays numerical identifiers on video frames and reformulates temporal grounding as frame-index reasoning for Vid-LLMs~\cite{wu2025number}. In contrast, our method introduces mask-conditioned visual markers, which provide object-centric evidence beyond global frame-level temporal cues and enable more consistent grounding of the queried entity over time. 

In addition, structure-aware inference further models compositional temporal structure, for example, by decomposing queries into ordered sub-events or organizing long videos into coherent parts, supporting training-free or weakly supervised localization with minimal model changes~\cite{zheng2024training,qu2024chatvtg,guo2024trace}. In contrast to methods that primarily rely on temporal annotations, frame-level indices, or query decomposition, our method targets a training-free setting and leverages mask-conditioned cues to enhance temporal localization and support more consistent object-centric tracking during grounding.

%-------------------------------------------------------------------------
\subsection{Mark for Video Understanding}
Recent work on visual prompting has shown that simple region cues, including graphic marks, numeric tags, and semantic masks, can guide vision-language models and multimodal LLMs to attend to specific regions and reduce spurious correlations~\cite{heo2025omni,yuan2025sa2va,nekrasov2025sa2va,munasinghe2025videoglamm,yuan2025videorefer}. This need for explicit and reliable visual evidence is also reflected in recent multimodal evaluation studies, where knowledge-centric benchmarks reveal that visually plausible outputs may still fail on reasoning-intensive image editing tasks~\cite{wu2026kris}. Representative approaches enable models to directly interpret overlaid marks for region-conditioned question answering, as in ViP-LLaVA~\cite{cai2024vip}, or to reference multiple regions via tag-based schemes. 

Beyond coarse marks, pixel-level prompts and mask-like cues have been explored to strengthen semantic localization~\cite{zou2023generalized,xia2024gsva,rasheed2024glamm,lai2024lisa}. For instance, CoLLaVO leverages a panoptic color map as a mark for object-centric understanding~\cite{lee2024collavo}, while Omni-RGPT extends token marks to videos with temporally consistent marks~\cite{heo2025omni}. More broadly, recent work on generative model controllability also highlights the importance of intervening on target concepts while preserving non-target visual concepts~\cite{wu2025unlearning}, which is related in spirit to our use of object-centric masks as controlled visual evidence. 

To make mark-conditioned reasoning practical at scale, promptable segmentation foundation models such as SAM~\cite{kirillov2023segment} and SAM 2~\cite{ravi2024sam} facilitate region mask generation and tracking, and can be composed with open-world detectors for text-conditioned mask acquisition. Collectively, these lines of work highlight segmentation as a form of structured visual evidence for video understanding, a perspective we adopt in our design.

\begin{figure*}[t]
\centering
\includegraphics[width=1.0\textwidth]{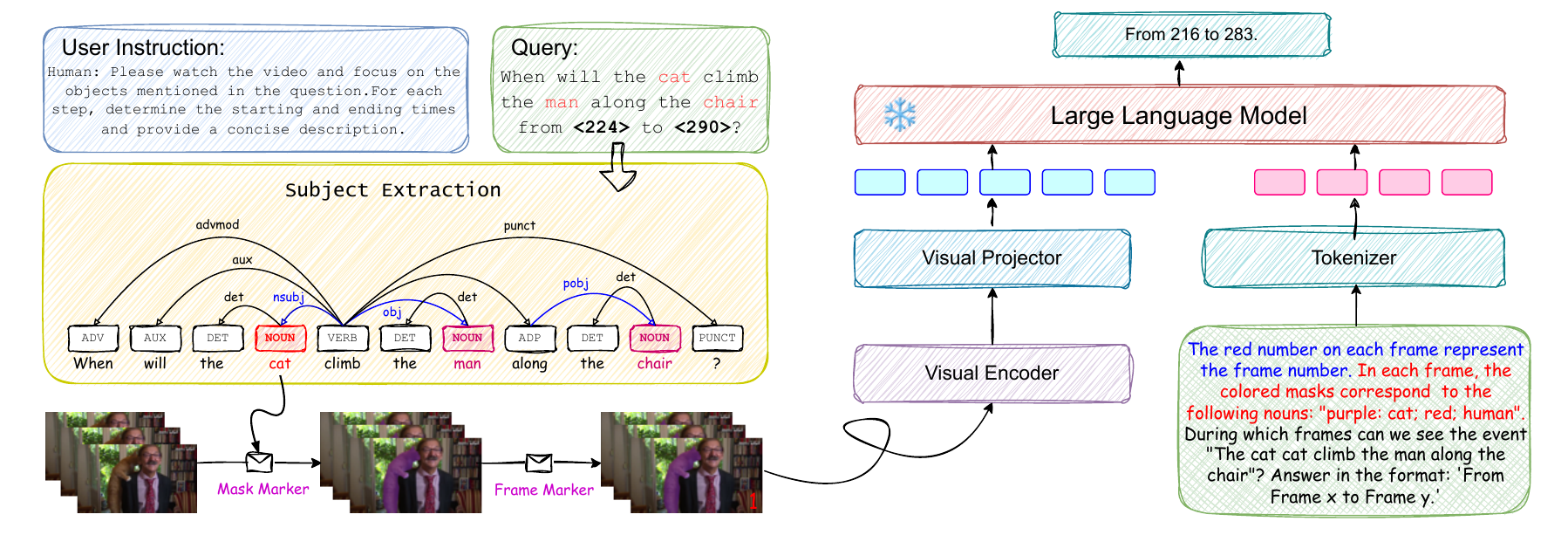}
\caption{Given a natural-language query, the system first performs syntactic parsing to extract subjects and relations, then injects subject masks and frame index markers into the video. The annotated video features are encoded and projected into a LLM together with tokenized text, enabling cross-modal reasoning and producing the predicted temporal interval of the queried event.}
\label{fig:main}
\end{figure*}
% Method (ICML-style LaTeX)
% -----------------------------
\section{Method}
\label{sec:method}

\subsection{Overview}
\label{sec:method:overview}

\MarkIt{} (Figure \ref{fig:main}) is a training-free markerization operator that rewrites the input video into a marked video, making it easier for Vid-LLMs to perform temporal grounding. The key idea is to externalize both (i) \emph{temporal reference} and (ii) \emph{query-relevant visual correspondence} as explicit, readable cues in the visual stream, without modifying any Vid-LLM weights.

Formally, for a given video $V=\{I_t\}_{t=1}^{T}$ with $T$ frames $I_t$ and a query $q$,  we define a transformation
\begin{equation}
\tilde{V} = \{\tilde{I}_t\}_{t=1}^{T} = \Phi(V, q),
\label{eq:phi_def}
\end{equation}
and obtain task outputs by
\begin{equation}
\hat{y} = \mathcal{F}(\tilde{V}, q; \pi) = \mathcal{F}(\Phi(V,q), q; \pi),
\label{eq:readout}
\end{equation}
where $\mathcal{F}$ is any VTG-capable Vid-LLM with its standard instruction $\pi$.
For moment retrieval, $\hat{y}$ corresponds to temporal boundaries $[t_s,t_e]$; for highlight detection,
$\hat{y}$ corresponds to query-relevance scores/rankings over clips or frames.
\MarkIt{} does not modify the weights of $\mathcal{F}$ (or any auxiliary component); it only changes the input video representation.

\noindent\textbf{Factorizing $\Phi$: Q2M-Bridge + markerization.}
We factor $\Phi$ into two stages: a \emph{Q2M-Bridge} and a rendering stage that overlays semantic markers and a persistent frame index on each frame (detailed in Sections below):
\begin{equation}
\begin{aligned}
& \{\mathcal{M}_t^{\mathrm{sem}}\}_{t=1}^{T}
= \QTwoM(V, q; K_{\max}) \qquad \text{(Q2M-Bridge)},\\
& \mathcal{M}_t^{\mathrm{idx}}
= \{(R^{\mathrm{idx}}, \mathrm{Index}(t))\}, \qquad t\in[1,T],\\
& \mathcal{M}_t
= \mathcal{M}_t^{\mathrm{sem}} \cup \mathcal{M}_t^{\mathrm{idx}},\\
& \tilde{I}_t
= \Render(I_t, \mathcal{M}_t; s), \qquad \tilde{V}=\{\tilde{I}_t\}_{t=1}^{T}.
\end{aligned}
\label{eq:pipeline}
\end{equation}
Here $\QTwoM$ denotes Q2M-Bridge, which maps $(V,q)$ to per-frame \emph{query-conditioned instance masks} and
associates them with canonical subject tags to form semantic markers $\{\mathcal{M}_t^{\mathrm{sem}}\}_{t=1}^{T}$ under budget $K_{\max}$ (Section~\ref{sec:method:q2m}). $\mathrm{Index}(t)$ produces a per-frame numeric identifier, $R^{\mathrm{idx}}$ is a fixed anchor region reserved for the frame index, and $\Render$ overlays all markers with a global style configuration $s$.

\subsection{Frame Index Markers}
\label{sec:method:index}

In addition to the semantic markers produced by Q2M-Bridge, \MarkIt{} assigns each frame $t$ a unique index text $\nu_t$
for explicit temporal reference. We define
\begin{equation}
\nu_t = \mathrm{Index}(t),
\label{eq:frame_index}
\end{equation}
where $\mathrm{Index}(\cdot)$ converts the integer frame id to a short text string (e.g., ``1'', ``2'', $\ldots$).
We render the index at a fixed anchor region $R^{\mathrm{idx}}\in\{0,1\}^{H\times W}$ (e.g., a small corner patch),
forming the index marker set
\begin{equation}
\mathcal{M}_t^{\mathrm{idx}}=\{(R^{\mathrm{idx}}, \nu_t)\}.
\label{eq:index_marker}
\end{equation}

\subsection{Visual Markers}
\label{sec:method:marker}

\vspace{3pt}\noindent\textbf{Marker representation.}
We represent any overlaid cue as a \emph{marker} $(R,\tau)$, where
$R\in\{0,1\}^{H\times W}$ denotes a region mask and $\tau$ is a short text string.

\vspace{3pt}\noindent\textbf{Per-frame marker set.}
For each frame $I_t$, \MarkIt{} renders the union of (i) semantic instance markers $\mathcal{M}_t^{\mathrm{sem}}$ and
(ii) a frame-index marker $\mathcal{M}_t^{\mathrm{idx}}$, i.e.,
\begin{equation}
\mathcal{M}_t = \mathcal{M}_t^{\mathrm{sem}} \cup \mathcal{M}_t^{\mathrm{idx}}.
\label{eq:marker_union}
\end{equation}

\vspace{0pt}\noindent\textbf{Rendering.}
Given a global style configuration $s$, the marked frame is obtained by
\begin{equation}
\tilde{I}_t = \Render(I_t, \mathcal{M}_t; s).
\label{eq:render}
\end{equation}

\subsection{Q2M-Bridge: Query-to-Mask Grounding}
\label{sec:method:q2m}

Q2M-Bridge is an annotation-free bridge that maps a natural-language query to query-conditioned instance masks, producing per-frame semantic marker sets $\{\mathcal{M}_t^{\mathrm{sem}}\}_{t=1}^{T}$ for \MarkIt{} to render. Concretely, Q2M-Bridge is implemented by composing (i) subject-tag proposal and (ii) text-conditioned open-vocabulary segmentation (Section~\ref{sec:method:proposal} and Section~\ref{sec:method:grounding}):
\begin{equation}
\begin{aligned}
& \mathcal{T} \!=\! [\tau_1,\ldots,\tau_{K_q}] \!=\! \mathcal{E}(q;K_{\max}), \ \ \   1 \le \! K_q \! \le K_{\max},\\
& \mathcal{R}_{t,i} \!=\! \mathcal{S}(I_t,\tau_i) \!=\! \{R^{(j)}_{t,i}\}_{j=1}^{J_{t,i}},  t\!\in\![1,T], \ \ \  i\!\in\![1,K_q],\\
& \mathcal{M}_t^{\mathrm{sem}} = \bigcup_{i=1}^{K_q}\ \{(R,\tau_i)\mid R \in \mathcal{R}_{t,i}\}, \ \ \ t\in[1,T],\\
& \{\mathcal{M}_t^{\mathrm{sem}}\}_{t=1}^{T} = \QTwoM(V,q;K_{\max}).
\end{aligned}
\label{eq:q2m}
\end{equation}
Here $\mathcal{E}$ is a prompted language model used as a structured subject-tag extractor under budget $K_{\max}$, and $\mathcal{S}$ is a text-conditioned open-vocabulary segmentation model that returns a set of instance masks for each tag.

\subsubsection{Subject Tag Proposal}
\label{sec:method:proposal}

We extract a small set of \emph{normalized visual subject classes} from the query $q$ as semantic anchors for grounding.
Let $\mathcal{T}=[\tau_1,\ldots,\tau_{K_q}]$ denote the extracted tags, where each $\tau_i$ is a lowercase, singularized class noun (e.g., \texttt{person}, \texttt{dog}, \texttt{car}). We cap the number of tags by a fixed budget $K_{\max}$ (in our implementation, $K_{\max}=3$).

\vspace{3pt}\noindent\textbf{Rule specification.}
Conceptually, the extraction operator $\mathcal{E}$ follows the specification
\begin{equation}
\begin{aligned}
\mathcal{T}
&=
\mathrm{FB}\!\left(
\mathrm{Dedup}_{K_{\max}}\!\left(
\mathrm{SC}\!\left(
\mathrm{Norm}\!\left(\mathrm{Subj}(q)\right)
\right)\right)\right),\\
&\qquad 1 \le K_q \le K_{\max},
\end{aligned}
\label{eq:subject_rules}
\end{equation}
where:
(i) $\mathrm{Subj}(q)$ extracts the grammatical subject(s) of the main action in $q$,
splitting coordinated subjects when necessary;
(ii) $\mathrm{Norm}(\cdot)$ normalizes each subject into a visually detectable noun class
by removing modifiers, singularizing plurals, and mapping human-related expressions to \texttt{person};
(iii) $\mathrm{SC}(\cdot)$ filters out non-subject entities such as objects, tools, or locations;
(iv) $\mathrm{Dedup}_{K_{\max}}(\cdot)$ removes duplicates and enforces a maximum of $K_{\max}$ tags;
and (v) $\mathrm{FB}(\cdot)$ ensures a non-empty output by falling back to a default subject when needed.

\vspace{3pt}\noindent\textbf{Implementation via prompted language model.}
We implement $\mathcal{E}$ using a prompted language model as a deterministic extractor:
\begin{equation}
\mathcal{T} = \mathcal{E}(q;K_{\max}) \;=\;
\Parse\!\left(\mathrm{LM}(\psi(q,K_{\max}))\right),
\label{eq:lm_extract}
\end{equation}
where: $\psi(q,K_{\max})$ is a fixed prompt template encoding Eq.~\eqref{eq:subject_rules}; $\mathrm{LM}(\cdot)$ returns \emph{only} a comma-separated list in lowercase; and
$\Parse(\cdot)$ splits by commas, trims whitespace, drops empty entries, and enforces the cap $K_{\max}$.

\subsubsection{Mask Grounding via Text-Conditioned Open-Vocabulary Segmentation}
\label{sec:method:grounding}

Given a subject tag $\tau_i\in\mathcal{T}$, we ground it to a set of instance masks on each frame
using a frozen text-conditioned open-vocabulary segmentation model $\mathcal{S}$:
\begin{equation}
\mathcal{R}_{t,i} = \mathcal{S}(I_t,\tau_i) = \{R^{(j)}_{t,i}\}_{j=1}^{J_{t,i}},
\label{eq:segmentation}
\end{equation}
where $R^{(j)}_{t,i}\in\{0,1\}^{H\times W}$ denotes the $j$-th grounded instance for tag $\tau_i$ on frame $t$.

\vspace{3pt}\noindent\textbf{Recall-first grounding.}
We retain all masks in $\mathcal{R}_{t,i}$ without additional confidence-based pruning or top-$K$ selection.
This follows our principle that redundant markers are acceptable while missing the correct region is not.

\vspace{3pt}\noindent\textbf{Aggregation into semantic markers.}
The per-frame semantic marker set is obtained by associating each retained mask with its tag:
\begin{equation}
\mathcal{M}_t^{\mathrm{sem}}
=
\bigcup_{i=1}^{K_q}\ \{(R,\tau_i)\mid R \in \mathcal{R}_{t,i}\},
\label{eq:sem_markers_from_seg}
\end{equation}
which is later united with the frame-index marker set to form $\mathcal{M}_t$
(see Section~\ref{sec:method:marker}).

\subsection{Marker Rendering}
\label{sec:method:rendering}

Given the per-frame semantic markers $\mathcal{M}_t^{\mathrm{sem}}$ produced by Q2M-Bridge
(Section~\ref{sec:method:q2m}) and the frame-index marker set $\mathcal{M}_t^{\mathrm{idx}}$
defined in Eq.~\eqref{eq:index_marker}, \MarkIt{} forms the full marker set
$\mathcal{M}_t=\mathcal{M}_t^{\mathrm{sem}}\cup \mathcal{M}_t^{\mathrm{idx}}$
and renders the marked frame by
\begin{equation}
\tilde{I}_t = \Render(I_t, \mathcal{M}_t; s).
\label{eq:render_def}
\end{equation}

A key design in \MarkIt{} is the \emph{rendering order}: we first render region cues (mask overlays and contours),
and then render all texts (both semantic tags and the frame index). Formally,
{\small
\begin{equation}
\tilde{I}_t
\!= \!
\Render_{\mathrm{text}}\!\Big(
\Render_{\mathrm{mask}}(I_t,\mathcal{M}_t^{\mathrm{sem}}; s_{\mathrm{mask}}),
\mathcal{M}_t; s_{\mathrm{text}}
\Big),
\label{eq:render_comp}
\end{equation}
}
\hspace{-0.07in}
where $s=(s_{\mathrm{mask}},s_{\mathrm{text}})$, and we denote by $I$ the intermediate canvas initialized as $I_t$
inside $\Render_{\mathrm{mask}}$.

\vspace{3pt}\noindent\textbf{Mask overlay.}
For each semantic marker $(R,\tau)\in\mathcal{M}_t^{\mathrm{sem}}$, we apply a translucent overlay with opacity $\alpha$:
\begin{equation}
I \leftarrow (1-\alpha R)\odot I + \alpha R \odot C_{\mathrm{mask}}(R,\tau),
\label{eq:mask_overlay}
\end{equation}
where $C_{\mathrm{mask}}(R,\tau)$ returns an RGB color for the marker (e.g., a fixed palette keyed by instance order;
single-color variants are a special case).

\vspace{3pt}\noindent\textbf{Contour enhancement.}
To make markers more salient, we additionally render a thin contour around each mask.
Let $\partial(R;w)\in\{0,1\}^{H\times W}$ denote a boundary operator that produces a contour band of width $w$
(e.g., via morphological gradient). We render the contour with opacity $\beta$:
\begin{equation}
I \leftarrow (1-\beta\,\partial(R;w))\odot I + \beta\,\partial(R;w)\odot C_{\mathrm{ctr}}(R,\tau),
\label{eq:mask_contour}
\end{equation}
where $C_{\mathrm{ctr}}(R,\tau)$ is the contour color (by default aligned with $C_{\mathrm{mask}}(R,\tau)$).

\vspace{3pt}\noindent\textbf{Fixed text placement.}
Unlike adaptive placement strategies that avoid overlap between text and masks, we adopt a fixed placement rule,
which empirically performs better in our setting.
For each marker $(R,\tau)\in\mathcal{M}_t$, we render text $\tau$ at a deterministic anchor
\begin{equation}
(x,y) = a(R),
\label{eq:text_anchor}
\end{equation}
where $a(\cdot)$ is a fixed anchor function (e.g., a chosen corner of the bounding box of $R$ with a constant offset).
We do not perform additional collision resolution between texts and masks.

\vspace{3pt}\noindent\textbf{Style configuration.}
We denote $s_{\mathrm{mask}}=(\alpha,\beta,w)$ and $s_{\mathrm{text}}=(\texttt{font},\texttt{scale},\texttt{offset})$.
Unless otherwise stated, a single default style is used across datasets and models.

\begin{table*}[t]
\centering
\footnotesize
\setlength{\tabcolsep}{4.5pt}
\renewcommand{\arraystretch}{1.15}
\caption{Quantitative comparison on Charades-STA, ActivityNet, and QVHighlights.}
\label{tab:numpro_vidllm}
\begin{tabular}{l|cccc|cccc|cc}
\toprule
\multirow{2}{*}{Model}
& \multicolumn{4}{c|}{Charades-STA}
& \multicolumn{4}{c|}{ActivityNet}
& \multicolumn{2}{c}{QVHighlights} \\
\cmidrule(lr){2-5}\cmidrule(lr){6-9}\cmidrule(lr){10-11}
& R@0.3 & R@0.5 & R@0.7 & mIoU
& R@0.3 & R@0.5 & R@0.7 & mIoU
& mAP & HIT@1 \\
\midrule
\midrule

\multicolumn{11}{c}{\textit{VTG-Tuned Vid-LLMs}} \\
\hline
GroundingGPT~\cite{groundinggpt}      & --   & 29.6 & 11.9 & --   & --   & --   & --   & --   & --   & --   \\
LITA~\cite{lita}                      & --   & --   & --   & --   & --   & 25.9 & --   & 28.6 & --   & --   \\
VTG-LLM~\cite{vtgllm}                 & 52.0 & 33.8 & 15.7 & --   & --   & --   & --   & --   & 16.5 & 33.5 \\
TimeChat~\cite{timechat}              & 47.7 & 22.9 & 12.5 & 30.6 & 30.2 & 16.9 & 8.2  & 21.8 & 14.5 & 23.9 \\
VTimeLLM~\cite{vtimellm}              & 51.0 & 27.5 & 11.4 & 31.2 & 44.0 & 27.8 & 14.3 & 30.4 & --   & --   \\
Momentor~\cite{momentor}              & 42.9 & 23.0 & 12.4 & 29.3 & 42.6 & 26.6 & 11.6 & 28.5 & 7.6  & --   \\
HawkEye~\cite{hawkeye}                & 50.6 & 31.4 & 14.5 & 33.7 & 49.1 & 29.3 & 10.7 & 32.7 & --   & --   \\
\midrule

\multicolumn{11}{c}{\textit{General Vid-LLMs}} \\
\hline

Qwen2-VL-7B~\cite{qwen2vl}
& 8.7  & 5.4  & 2.4  & 7.9
& 17.0 & 9.4  & 3.9  & 12.5
& 21.5 & 42.2 \\
\rowcolor{black!8}
\quad +MarkIt
& \textbf{62.3} & \textbf{38.8} & \textbf{16.6} & \textbf{41.1}
& \textbf{47.7} & \textbf{28.1} & \textbf{16.0} & \textbf{33.3}
& \textbf{24.1} & \textbf{44.1} \\
\hline

LLaVA-OV-7B~\cite{li2024llava}
& 23.3  & 5.7  & 3.1  & 10.0
& 17.7 & 8.0  & 4.2  & 13.3
& 18.1 & 41.0 \\
\rowcolor{black!8}
\quad +MarkIt
& \textbf{33.0} & \textbf{13.3} & \textbf{17.8} & \textbf{21.6}
& \textbf{37.4} & \textbf{17.4} & \textbf{9.0} & \textbf{22.4}
& \textbf{23.0} & \textbf{42.0} \\
\hline

InternVL2-8B~\cite{chen2024internvl}
& 21.0  & 9.1  & 1.6  & 19.2
& 10.2 & 6.1  & 4.1  & 17.1
& 19.4 & 33.2 \\
\rowcolor{black!8}
\quad +MarkIt
& \textbf{38.7} & \textbf{10.9} & \textbf{2.4} & \textbf{23.3}
& \textbf{20.3} & \textbf{10.0} & \textbf{5.2} & \textbf{24.6}
& \textbf{21.4} & \textbf{34.5} \\
\hline

LongVA-7B-DPO~\cite{longva}
& 22.6 & 10.1 & 2.2  & 14.6
& 11.8 & 5.3  & 1.9  & 8.2
& 14.2 & 20.4 \\

\quad +NumPro
& 27.2 & 10.3 & 2.9  & 18.9
& 20.1 & 10.8 & 5.4  & 15.2
& 15.3 & 24.3 \\

\rowcolor{black!8}
\quad +MarkIt
&  \textbf{32.6} & \textbf{11.6} & \textbf{3.0} & \textbf{21.8} 
& \textbf{22.9} & \textbf{12.0} & \textbf{5.7} & \textbf{17.0}
& \textbf{18.2} & \textbf{27.4} \\

\quad +NumPro-FT
& 63.8 & 42.0 & 20.6 & 41.4
& 55.6 & 37.5 & 20.6 & 38.8
& 25.0 & 37.2 \\
\rowcolor{black!8}
\quad +MarkIt-FT
& \textbf{65.1} & \textbf{43.2} & \textbf{21.8} & \textbf{43.9}
& \textbf{59.1} & \textbf{39.6} & \textbf{22.1} & \textbf{40.1}
& \textbf{25.9} & \textbf{39.1} \\
\bottomrule
\end{tabular}
\end{table*}

\subsection{Temporal Grounding with Marked Videos}
\label{sec:method:inference}

After markerization, we obtain the marked video $\tilde{V}=\{\tilde{I}_t\}_{t=1}^{T}$.
We then apply any VTG-capable video-language model $\mathcal{F}$ to produce task outputs
\begin{equation}
\hat{y} = \mathcal{F}(\tilde{V}, q; \pi),
\label{eq:readout2}
\end{equation}
where $\pi$ is the standard instruction/prompt used by $\mathcal{F}$. Note again that:
for moment retrieval, $\hat{y}$ corresponds to temporal boundaries $[t_s,t_e]$; for highlight detection,
$\hat{y}$ corresponds to query-relevance scores/rankings over clips or frames.
\MarkIt{} is training-free: it only rewrites the input representation and does not update any model weights.

% \subsection{Algorithm}
% \label{sec:method:algo}

% \begin{algorithm}[t]
% \caption{Training-free \MarkIt{} inference}
% \label{alg:markit}
% \begin{algorithmic}[1]
% \REQUIRE Video $V=\{I_t\}_{t=1}^{T}$, query $q$, tag budget $K_{\max}$,
% extractor $\mathcal{E}$, segmenter $\mathcal{S}$, VTG model $\mathcal{F}$, style $s$
% \ENSURE $[t_s,t_e]$

% \STATE $\mathcal{T}\leftarrow \mathcal{E}(q;K_{\max})$ \hfill // normalized subject classes
% \FOR{$t=1$ \textbf{to} $T$}
%   \STATE $\mathcal{M}_t^{\mathrm{sem}} \leftarrow \emptyset$
%   \FOR{each $\tau_i \in \mathcal{T}$}
%     \STATE $\mathcal{R}_{t,i} \leftarrow \mathcal{S}(I_t,\tau_i)$ \hfill // a set of masks
%     \STATE $\mathcal{M}_t^{\mathrm{sem}} \leftarrow \mathcal{M}_t^{\mathrm{sem}} \cup \{(R,\tau_i)\mid R\in\mathcal{R}_{t,i}\}$
%   \ENDFOR
%   \STATE $\nu_t \leftarrow \mathrm{Index}(t)$
%   \STATE $R_t^{\mathrm{idx}} \leftarrow \mathrm{MaskFromIndex}(\nu_t)$
%   \STATE $\mathcal{M}_t^{\mathrm{idx}} \leftarrow \{(R_t^{\mathrm{idx}}, \nu_t)\}$
%   \STATE $\mathcal{M}_t \leftarrow \mathcal{M}_t^{\mathrm{sem}} \cup \mathcal{M}_t^{\mathrm{idx}}$
%   \STATE $\tilde{I}_t \leftarrow \Render(I_t,\mathcal{M}_t;s)$
% \ENDFOR
% \STATE $\tilde{V}\leftarrow\{\tilde{I}_t\}_{t=1}^{T}$
% \STATE $[t_s,t_e] \leftarrow \mathcal{F}(\tilde{V},q;\pi)$
% \STATE \textbf{return} $[t_s,t_e]$
% \end{algorithmic}
% \end{algorithm}

% -----------------------------
% Experiments (ICML-style LaTeX)
% -----------------------------
\section{Experiments}
\label{sec:exp}

% Optional: list the evaluated Vid-LLMs (fill in your final paper).
\newcommand{\VidLLMs}{Model-A, Model-B, Model-C}

\subsection{Tasks, Datasets, and Metrics}
\label{sec:exp:task}
We evaluate \MarkIt{} on two standard \VTG{} tasks following prior work: Moment Retrieval and Highlight Detection.
For moment retrieval, we evaluate on Charades-STA~\cite{gao2017tall} and ActivityNet~\cite{caba2015activitynet}, where the model predicts the start and end temporal boundaries given a language query. We report $\mIoU$ and $\R{1}$ at IoU thresholds $\{0.3, 0.5, 0.7\}$.
For highlight detection, we employ QVHighlights~\cite{lei2021detecting}, where the model ranks clips/frames by query relevance. We report $\mAP$ and $\HIT$.

\subsection{Evaluation Protocols}
\label{sec:exp:protocol}

\vspace{3pt}\noindent\textbf{Training-free and SFT protocol.}
We evaluate \MarkIt{} under both training-free and SFT settings.
In the training-free setting, model weights are frozen and the only change is applying the markerization operator $\Phi(V,q)$ (Section~\ref{sec:method}) to rewrite the input video.
For SFT, we fine-tune LongVA-7B-DPO~\cite{zhang2024long} following the data split and hyperparameters of Number-IT~\cite{wu2025number}.
All methods are trained and evaluated under the same protocol, the difference remains whether the input videos are markerized by \MarkIt{}.

\vspace{3pt}\noindent\textbf{Prompt format and answer parsing.}
For moment retrieval, we use a fixed instruction template consistent across methods:
% \begin{quote}
% \small \texttt{During which frames can we see \{query\}?}
% \end{quote}
\noindent\texttt{\small During which frames can we see \{query\}?}
The model is required to output a span in the form \texttt{From x to y}.
We parse integers $(x,y)$ from the response and clamp them to the valid frame range if needed.

For highlight detection, we use a fixed instruction template that asks the model to output the highlight frame ids and their saliency scores. We parse the returned indices and scores and convert them to per-clip saliency predictions following the dataset protocol.

Prompt templates are given in Appendix \ref{app:prompts}.

\vspace{3pt}\noindent\textbf{Video preprocessing.}
We adopt the same video preprocessing pipeline and apply it consistently across all methods and settings to ensure fair comparison.

\begin{table}[t]
\centering
\caption{Ablation on subject tag proposal strategies ($\mathcal{E}$) on Charades-STA and ActivityNet.}
\label{tab:abl_tag_both}
\footnotesize
\setlength{\tabcolsep}{5pt}
\begin{tabular}{lcccc}
\toprule
Tag strategy & $\R0.3$ & $\R0.5$ & $\R0.7$ & $\mIoU$ \\
\midrule
\midrule
\multicolumn{5}{l}{\textbf{Charades-STA}} \\
\hline
No nouns        & 19.09 &  6.80 & 1.37 & 12.93 \\
All nouns       & 29.27 & 10.86 & 3.36 & 19.11 \\
Single noun     & 30.67 & \textbf{12.04} & \textbf{3.68} & 19.86 \\
Subject nouns   & \textbf{31.85} & 11.94 & 3.36 & \textbf{20.13} \\
\midrule
\multicolumn{5}{l}{\textbf{ActivityNet}} \\
\hline
No nouns        & 11.80 &  5.30 & 1.90 &  8.20 \\
All nouns       & 20.53 &  9.17 & 3.25 & 14.69 \\
Single noun     & 23.05 & 10.41 & 3.76 & 16.10 \\
Subject nouns   & \textbf{23.20} & \textbf{10.68} & \textbf{4.00} & \textbf{16.30} \\
\bottomrule
\end{tabular}
\end{table}

% \subsection{Implementation Details}
% \label{sec:exp:impl}

% \textbf{Marker budget.}
% We set the query-level tag budget to $K_{\max}=3$ by default unless otherwise stated.

% \textbf{Subject tag proposal.}
% We use the same subject-extraction prompt template for all datasets and models, and decode deterministically
% (e.g., greedy decoding and temperature $=0$) to ensure stable outputs.

% \textbf{Open-vocabulary grounding.}
% Given each tag, we apply YOLOE-Large~\cite{wang2025yoloe} as a text-conditioned open-vocabulary segmentation model
% to obtain a set of instance masks per frame, and we retain all returned masks (recall-first grounding).

% % Given each tag, we apply a text-conditioned open-vocabulary segmentation model to obtain a set of instance masks
% % per frame, and we retain all returned masks (recall-first grounding).

% \textbf{Rendering configuration.}
% Unless otherwise stated, we use a default rendering style $s$ that overlays each mask with opacity $\alpha$,
% adds a contour of width $w$ for saliency, and renders text after rendering the mask overlay (mask $\rightarrow$ text).
% Frame indices are rendered in a fixed anchor region, and semantic tag texts are placed by a deterministic rule.

\begin{table*}[t]
\centering
\caption{Ablation on subject-tag extraction (fill opacity $\alpha$, contour opacity $\beta$, contour width $w$, and color) on ActivityNet.}
\label{tab:abl_mask}
\footnotesize
\setlength{\tabcolsep}{12pt}
\begin{tabular}{lccccccc}
\toprule
Rendering style & $\alpha$ & $\beta$ & Contour ($w$) & $\R0.3$ & $\R0.5$ & $\R0.7$ & $\mIoU$ \\
\midrule
\midrule
No mask                   & --  & --  & --        & 11.80 &  5.30 & 1.90 &  8.20 \\
\midrule
Mask fill (palette)                  & 0.2 & --  & off       & 22.81 & 10.45 & 3.96 & 15.99 \\
Mask fill (palette)                  & 0.3 & --  & off       & 23.10 & 10.63 & 4.00 & 16.20 \\
Mask fill (palette)                  & 0.5 & --  & off       &  9.29 &  4.20 & 1.43 &  6.56 \\
Mask fill (all-red)                  & 0.3 & --  & off       & 22.64 & 10.62 & \textbf{4.09} & 16.10 \\
\midrule
Mask + contour (palette)             & 0.3 & 1.0 & $w=2$     & --    & 10.77 & 3.95 & 16.32 \\
Mask + contour (palette)             & 0.3 & 1.0 & $w=3$     & \textbf{23.42} & \textbf{10.70} & 3.96 & \textbf{16.35} \\
Mask + contour (palette)             & 0.3 & 1.0 & $w=5$     & 23.20 & 10.52 & 3.84 & 16.23 \\
\bottomrule
\end{tabular}
\end{table*}

\subsection{Implementation Details}
\label{sec:exp:impl}

\vspace{0pt}\noindent\textbf{Q2M-Bridge budget.}
%✅
Unless otherwise stated, we set the tag budget in $\QTwoM(V,q;K_{\max})$ to $K_{\max}=3$.

\vspace{3pt}\noindent\textbf{Subject tag proposal.}
We implement $\mathcal{E}$ (Section~\ref{sec:method:proposal}) with a fixed prompt template and deterministic decoding
(greedy, temperature $=0$), and use the same extractor across all datasets and Vid-LLM backbones.

\vspace{3pt}\noindent\textbf{Mask grounding.}
%✅
We instantiate the text-conditioned segmentation model with
YOLOE-Large~\cite{wang2025yoloe}. For each tag $\tau_i\in\mathcal{T}$ and each frame $I_t$,
$\mathcal{S}(I_t,\tau_i)$ outputs a set of instance masks $\mathcal{R}_{t,i}$, all of which are retained (recall-first).

\vspace{3pt}\noindent\textbf{Markerization with index.}
%✅
We generate the marked video by $\tilde{I}_t=\Render(I_t,\mathcal{M}_t;s)$ (Section~\ref{sec:method:rendering})
using a single default style $s=(s_{\mathrm{mask}},s_{\mathrm{text}})$ unless otherwise stated.
We follow the fixed order \texttt{mask}$\rightarrow$\texttt{text} (Eq.~\eqref{eq:render_comp}),
place semantic tag texts with the deterministic anchor $a(R)$, and render the frame index at the fixed region $R^{\mathrm{idx}}$.

\subsection{Main Results}
\label{sec:exp:main}
Table~\ref{tab:numpro_vidllm} reports quantitative results on moment retrieval
(Charades-STA and ActivityNet) and highlight detection (QVHighlights),
covering both VTG-tuned models and general-purpose Vid-LLMs.

\vspace{3pt}\noindent\textbf{Moment retrieval.}
We first evaluate \MarkIt{} in a training-free setting on general Vid-LLMs. In Table~\ref{tab:numpro_vidllm}, vanilla Vid-LLMs exhibit limited temporal grounding ability.

In contrast, \MarkIt{} provides substantially larger improvements. On Charades-STA, \MarkIt{} improves LongVA to 21.8 $\mIoU$, with consistent recall gains across IoU thresholds (e.g., $\R0.3$ from 22.6 to 32.6). On ActivityNet, $\mIoU$ is further increased from 15.2 to 17.0,
with improved high-overlap recall. Importantly, these gains generalize across different Vid-LLM backbones. Applying \MarkIt{} to Qwen2-VL-7B boosts $\mIoU$ from 7.9 to 41.1 on Charades-STA and from 12.5 to 33.3 on ActivityNet, outperforming several VTG-tuned methods.
Similar trends are observed on LLaVA-OV-7B and InternVL2-8B, demonstrating that semantic region-aware markers can effectively unlock temporal grounding in general-purpose Vid-LLMs without task-specific training.

When SFT is available, \MarkIt{} remains beneficial. On Charades-STA, \MarkIt{-FT} improves $\mIoU$ from 14.6 to 43.9, and on ActivityNet from 8.2 to 40.1, indicating that explicit semantic markers complement end-to-end optimization.

\vspace{3pt}\noindent\textbf{Highlight detection.}
On QVHighlights, the vanilla LongVA baseline achieves 14.2 $\mAP$ and 20.4 $\HIT$,
which are modestly improved by \NumPro{}.
\MarkIt{} yields larger gains, reaching 18.2 $\mAP$ and 27.4 $\HIT$
in the training-free setting.

After fine-tuning, \MarkIt{-FT} achieves the best overall performance
with 25.9 $\mAP$ and 39.1 $\HIT$,
further confirming the effectiveness of semantic region-aware visual prompting
for highlight detection.

\subsection{Ablation Studies}
\label{sec:exp:ablation}

We conduct extensive ablations to understand which design choices contribute most to \MarkIt{}.

\vspace{3pt}\noindent{\textbf{Ablation on  Q2M-Bridge.}}
\label{sec:exp:abl_tag}
Following the design in Section~3, we use Qwen-7B as the subject-tag extraction operator,
which determines the query-conditioned semantic anchors used by Q2M-Bridge.
We compare four strategies:
\emph{No nouns} (Q2M-Bridge disabled),
\emph{All nouns} (all noun phrases),
\emph{Single noun} (most salient noun),
and \emph{Subject nouns (ours)}, which extracts grammatical subjects
as defined in Eq.~\eqref{eq:subject_rules}.
Queries may contain multiple subject nouns, all of which are retained.

Table~\ref{tab:abl_tag_both} reports results on Charades-STA and ActivityNet.
Disabling $\mathcal{E}$ causes a clear performance drop, highlighting the importance
of explicit semantic anchors.
While extracting all nouns improves over this baseline, it introduces noise.
More focused strategies perform better, with subject-noun extraction achieving
the best overall trade-off, yielding the highest $\mIoU$ and $\R0.3$.
These results validate the subject-centric design of Q2M-Bridge.

\begin{table}[t]
\centering
\caption{Ablation on NumPro-style frame-index overlay (placement, font size, and color) on ActivityNet with the best \MarkIt{} mask setting fixed.}
\label{tab:abl_text_activitynet}
\footnotesize
\setlength{\tabcolsep}{4pt}
\resizebox{\columnwidth}{!}{
\begin{tabular}{lllcccc}
\toprule
Size & Color & Position & $\R0.3$ & $\R0.5$ & $\R0.7$ & $\mIoU$ \\
\midrule
\midrule
\multicolumn{7}{l}{\textit{Position ablation}} \\
\hline
40 & Black & Top Left      & 21.90 & 11.20 & \textbf{5.40} & 16.50 \\
40 & Black & Top Right     & 21.70 & 11.10 & 5.30 & 16.40 \\
40 & Black & Center        & 21.30 & 10.80 & 5.10 & 16.10 \\
40 & Black & Bottom Left   & 21.60 & 11.00 & 5.25 & 16.30 \\
40 & Black & Bottom Right  & \textbf{21.93} & \textbf{11.57} & 4.95 & \textbf{16.58} \\
40 & Black & Find region   & 20.87 & 10.97 & 4.24 & 16.26 \\
\midrule
\multicolumn{7}{l}{\textit{Size ablation}} \\
\hline
20 & Black & Bottom Right  & 20.10 & 10.10 & 4.70 & 15.10 \\
30 & Black & Bottom Right  & 20.60 & 10.40 & 4.90 & 15.40 \\
38 & Black & Bottom Right  & \textbf{22.87} & \textbf{11.97} & \textbf{5.74} & \textbf{16.96} \\
40 & Black & Bottom Right  & 21.93 & 11.57 & 4.95 & 16.58 \\
\midrule
\multicolumn{7}{l}{\textit{Color ablation}} \\
\hline
38 & Black & Bottom Right  & \textbf{22.87} & \textbf{11.97} & \textbf{5.74} & \textbf{16.96} \\
38 & Red   & Bottom Right  & 20.70 & 10.35 & 4.85 & 15.40 \\
38 & Blue  & Bottom Right  & 20.80 & 10.40 & 4.88 & 15.45 \\
\bottomrule
\end{tabular}
}
\end{table}

\begin{figure*}[t]
\centering
\includegraphics[width=1.0\textwidth]{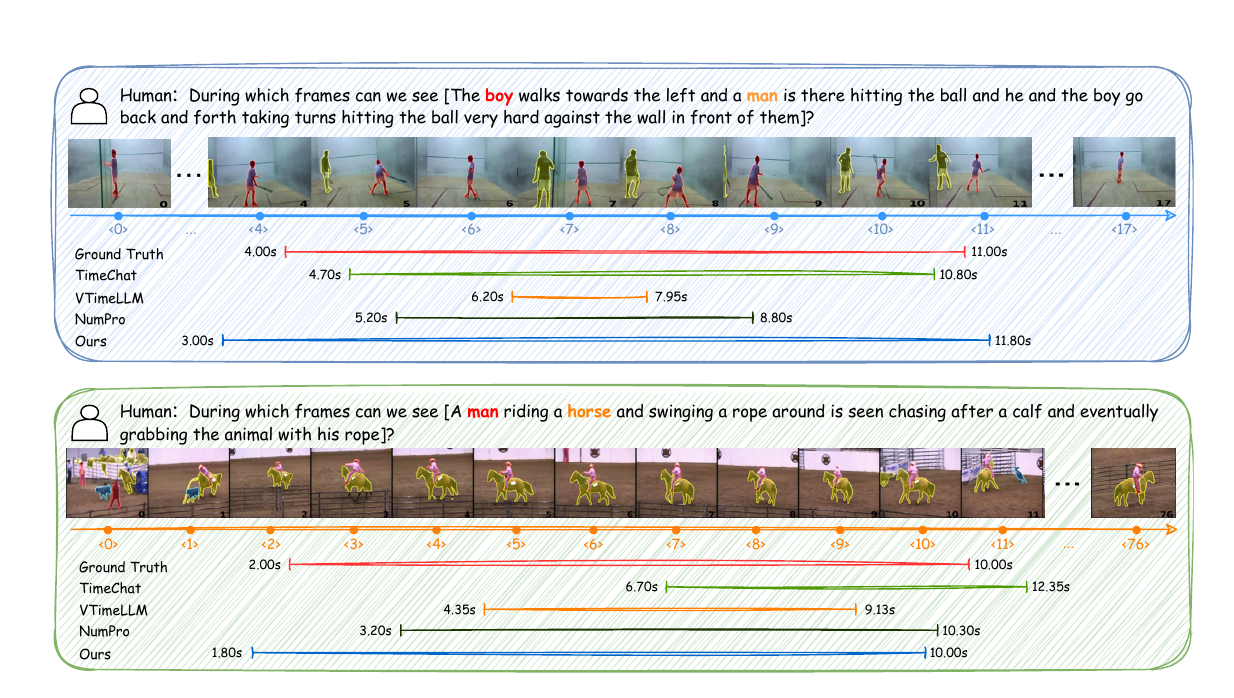}
\caption{Qualitative comparison on ActivityNet. Predicted spans vs.\ ground truth.
\MarkIt{} more reliably delineates event boundaries than TimeChat~\cite{ren2024timechat} and
VTimeLLM~\cite{huang2024vtimellm} in challenging scenarios.}
\label{fig:qual}
\end{figure*}

\vspace{3pt}\noindent{\textbf{Ablation on mask rendering.}}
\label{sec:exp:abl_mask}
We ablate the mask-rendering component $\Render_{\mathrm{mask}}$ to examine how visual saliency affects temporal grounding, focusing on the fill opacity $\alpha$, contour opacity $\beta$, and contour width $w$.
For \textit{mask fill (palette)}, instance ids are visualized with a fixed discrete color mapping
(background unfilled; foreground ids rendered as $1$=red, $2$=yellow, $3$=blue, $4$=green) at opacity $\alpha$.
For \textit{mask fill (all-red)}, all foreground ids share the same red fill with opacity $\alpha$.
For \textit{contour}, mask boundaries are overlaid using the same id-to-color mapping,
with opacity $\beta$ and pixel width $w$.

Table~\ref{tab:abl_mask} reports representative settings on ActivityNet. Removing the mask yields poor performance ($\mIoU=8.20$). Among fill-only variants, moderate opacity performs best: palette fill with $\alpha=0.3$ reaches $\mIoU=16.20$, while high opacity ($\alpha=0.5$) severely degrades results ($\mIoU=6.56$). Adding contours further improves performance, with a medium width ($w=3$) achieving the best overall trade-off ($\R0.3=23.42$, $\R0.5=10.70$, $\mIoU=16.35$), indicating that moderate visual emphasis is most effective.

\vspace{3pt}\noindent\textbf{Ablation on frame-index rendering.}
\label{sec:exp:abl_text}
To further strengthen temporal grounding, we combine \MarkIt{} with NumPro-style frame-index overlays, so that each frame is annotated with both semantic region markers (ours) and explicit temporal indices (NumPro). 

Table~\ref{tab:abl_text_activitynet} summarizes results on ActivityNet, while the corresponding Charades-STA study is
provided in Appendix Table~\ref{tab:abl_text_charades}.
As shown in Table~\ref{tab:abl_text_activitynet}, a clean and unobtrusive overlay is preferred:
fixed-corner placement (NumPro default) and placing indices in an empty region perform similarly, whereas center placement is consistently worse.
A medium font size (38) achieves the best overall results on ActivityNet ($\mIoU=16.96$, $\R0.7=5.74$),
and black text performs best among color choices.

Further ablation studies are given in Appendix \ref{app:ablation}.

\subsection{Qualitative Results}
\label{sec:exp:qual}

Figure~\ref{fig:qual} compares \MarkIt{} with strong baselines on ActivityNet.
In the first example, \MarkIt{} produces temporal predictions that align more closely with the ground-truth boundaries.
In the second example, \MarkIt{} better focuses on the target event and avoids including irrelevant segments, whereas other methods may exhibit certain deviations, such as predicting overly short temporal spans.

\vspace{0.08in}

\section{Conclusion}
\label{sec:conclusion}
We introduce \MarkIt{}, a training-free markerization operator for video temporal grounding that injects frame indices and subject-aware instance markers to make temporal reference and visual correspondence explicit. Without modifying model weights, \MarkIt{} enables Vid-LLMs to produce more reliable temporal predictions. Experiments on moment retrieval and highlight detection show consistent improvements across multiple Vid-LLM backbones, validating the effectiveness of our design.

\vspace{0.08in}

\bibliographystyle{IEEEtran}
\bibliography{references}

%%%%%%%%%%%%%%%%%%%%%%%%%%%%%%%%%%%%%%%%%%%%%%%%%%%%%%%%%%%%%%%%%%%%%%%%%%%%%%%
%%%%%%%%%%%%%%%%%%%%%%%%%%%%%%%%%%%%%%%%%%%%%%%%%%%%%%%%%%%%%%%%%%%%%%%%%%%%%%%
% APPENDIX
%%%%%%%%%%%%%%%%%%%%%%%%%%%%%%%%%%%%%%%%%%%%%%%%%%%%%%%%%%%%%%%%%%%%%%%%%%%%%%%
%%%%%%%%%%%%%%%%%%%%%%%%%%%%%%%%%%%%%%%%%%%%%%%%%%%%%%%%%%%%%%%%%%%%%%%%%%%%%%%
\newpage
\appendix

\subsection{Prompt Templates: System Message for Normalized Subject Extraction}
\label{app:prompts}

\begin{Verbatim}[fontsize=\footnotesize,breaklines]
You are an NLP tool for extracting normalized visual subjects for open-vocabulary object detection.
The input is an English sentence describing an action in a video.
Your job is to return ONLY the grammatical subject(s), normalized into simple noun classes.

Important:
- The input is always in English. Do NOT translate anything. Do NOT output any Chinese.
- Only output the final result as a comma-separated list in lowercase.
- Do NOT output any explanations, steps, or extra words.

Subject identification (Stage 1):
- Find who or what performs the main action in the sentence (the grammatical subject).
- If there are several subjects joined by 'and' or commas (e.g. 'I, my dog and my cat'), treat each as a separate subject.
- If the subject is a vague pronoun referring to people (e.g. 'some', 'someone', 'everyone', 'others'), treat it as a human subject.
- Ignore nouns that are NOT subjects (objects, tools, locations, goals, etc.).

Normalization rules (Stage 2):
- Keep only entities that could be visually detected in a frame (people, animals, objects, visible on-screen text like credits).
- Remove determiners and possessives: 'the man', 'a woman', 'my dog' -> 'man', 'woman', 'dog'.
- Remove quantity words: 'two men', 'three cats', 'a group of people' -> 'man', 'cat', 'person'.
- For human pronouns ('I', 'you', 'he', 'she', 'we', 'they') and vague human pronouns ('some', 'someone', 'everyone', 'others', 'anyone'), normalize to 'person'.
- Singularize plurals: 'men' -> 'man', 'cats' -> 'cat', 'people' -> 'person'.
- Drop descriptive modifiers and keep the core class noun: 'man wearing athletic gear' -> 'man'.
- If several subjects normalize to the same word, keep only one and preserve the order.

Self-check and fallback (Stage 3):
- Silently check each candidate: if it is being acted ON (object), used as a tool, or only appears inside a prepositional phrase, REMOVE it.
- This is video data: there is always some visible entity. You MUST always output at least one subject.
- If no clear subject remains after self-check, choose the most likely main visible entity:
  * First, prefer 'person' if people are implied.
  * Otherwise, choose the first concrete noun in the sentence (e.g. 'ball', 'car', 'credits').

Output:
- Output at most 3 normalized subjects.
- Output format: a comma-separated list, lowercase, e.g. 'person, dog'.
- No explanations, no JSON, no extra tokens.

Examples:
Sentence: "Two men both dressed in athletic gear are standing and talking in an indoor weight lifting gym filled with other equipment." -> man
Sentence: "One man is holding onto a rope attached to a machine, and the other man instructs him to bend down on his left knee while still holding onto the rope and showing the man how to have proper form." -> man
Sentence: "The man then instructs the man holding the rope to pull the row down a few times and he's talking the whole time." -> man
Sentence: "I, my dog and my cat are running together in the park." -> person, dog, cat
Sentence: "The credits of the clip are shown." -> credits
Sentence: "... and some are in wheel chairs." -> person
Sentence: "A man is holding a rope in a gym." -> man  (do NOT output 'rope' or 'gym')
Sentence: "A man pushes a woman in a wheel chair across the room." -> man  (do NOT output 'woman' or 'wheel chair' or 'room')
If everything is unclear, still choose the most likely main entity and output one subject.
\end{Verbatim}

\subsection{Additional Analysis on Frame-index Rendering (Charades-STA)}
\label{sec:exp:abl_frame_index}

Table~\ref{tab:abl_text_charades} reports the corresponding ablation study on Charades-STA.
Overall trends are consistent with those observed on ActivityNet.
Specifically, fixed-corner placements yield comparable performance, while center placement remains suboptimal,
suggesting that intrusive overlays may interfere with visual grounding.
In terms of font size, a medium scale (30–38) provides a favorable trade-off between readability and visual clutter,
leading to higher $\mIoU$ and $\R0.7$.
Finally, black text consistently outperforms colored alternatives, indicating that high-contrast yet neutral rendering
is preferable for preserving both semantic cues and temporal clarity.

\begin{table}[h]
\centering
\caption{Ablation on NumPro-style frame-index overlay (placement, font size, and color) on Charades-STA with the best \MarkIt{} mask setting fixed.}
\label{tab:abl_text_charades}
\small
\setlength{\tabcolsep}{5pt}
\begin{tabular}{lllcccc}
\toprule
Size & Color & Position & $\R0.3$ & $\R0.5$ & $\R0.7$ & $\mIoU$ \\
\midrule
\midrule
\multicolumn{7}{l}{\textit{Position ablation}} \\
\hline
40 & Black & Top Left      & 31.05 &  9.53 & 1.85 & 20.68 \\
40 & Black & Top Right     & 31.21 &  9.49 & 2.12 & 20.94 \\
40 & Black & Center        & 30.65 &  8.84 & 1.51 & 21.05 \\
40 & Black & Bottom Left   & 30.94 & 11.13 & 2.31 & 21.07 \\
40 & Black & Bottom Right  & 31.67 &  9.60 & 2.15 & 21.04 \\
40 & Black & Find region   & 30.67 & 10.11 & 2.07 & 20.80 \\
\midrule
\multicolumn{7}{l}{\textit{Size ablation}} \\
\hline
20 & Black & Bottom Right  & 29.00 &  9.67 & 2.00 & 20.83 \\
30 & Black & Bottom Right  & \textbf{33.33} &  8.67 & \textbf{3.33} & \textbf{23.10} \\
38 & Black & Bottom Right  & 32.53 & \textbf{11.64} & 2.80 & 21.82 \\
40 & Black & Bottom Right  & 31.67 &  9.60 & 2.15 & 21.04 \\
\midrule
\multicolumn{7}{l}{\textit{Color ablation}} \\
\hline
38 & Black & Bottom Right  & 32.53 & \textbf{11.64} & \textbf{2.80} & 21.82 \\
38 & Red   & Bottom Right  & 30.54 & 10.16 & 1.96 & 20.86 \\
38 & Blue  & Bottom Right  & 31.61 & 11.40 & 2.47 & 21.71 \\
\bottomrule
\end{tabular}
\end{table}

\subsection{Ablation on Color Parameterization}
\label{sec:exp:abl_color}
Since visual attention models treat color contrast as a primary cue and integrate it with
intensity and orientation to form saliency, we examine whether different color
parameterizations of \MarkIt{} masks affect how reliably region markers attract attention
during temporal grounding.

Table~\ref{tab:abl_color} reports the effect of different color parameterizations on ActivityNet.
The default RGB palette achieves the best overall performance
($\mIoU=16.35$, $\R0.5=10.70$),
while using high-saturation/value HSV colors yields comparable but slightly worse results.
In contrast, low-saturation/value palettes lead to a noticeable drop
($\mIoU=15.48$),
suggesting that sufficient color contrast is important for reliable region highlighting,
but excessively vivid colors do not provide additional benefits.

\begin{table}[h]
\centering
\caption{Ablation on mask color parameterization on ActivityNet (fixed $\alpha=0.3$, $\beta=1.0$, $w=3$).}
\label{tab:abl_color}
\footnotesize
\setlength{\tabcolsep}{4pt}
\begin{tabular}{lcccc}
\toprule
Color config & $\R0.3$ & $\R0.5$ & $\R0.7$ & $\mIoU$ \\
\midrule
RGB palette (ours)       & \textbf{23.42} & \textbf{10.70} & \textbf{3.96} & \textbf{16.35} \\
HSV palette (high S/V)   & 23.35 & 10.39 & 3.64 & 16.31 \\
HSV palette (low S/V)    & 22.16 &  9.48 & 3.34 & 15.48 \\
\bottomrule
\end{tabular}
\end{table}

\end{document}